\documentclass[doublecol]{epl2} 
\usepackage{graphicx}
\usepackage{textcomp}
\usepackage{amsmath,amssymb}
\usepackage{color}
\usepackage{graphicx}
\usepackage{epsfig}
\usepackage{subfigure}
\usepackage{times}
\usepackage{color}
\usepackage{epstopdf}
\usepackage{amsthm}
\usepackage{bm}
\usepackage{amsmath}
\usepackage{amssymb}

 \title{Qubit(s) transfer in helical spin chains}
\author{ Harshit Verma\inst{1}, L.~Chotorlishvili\inst{2}, J.~Berakdar\inst{2} \and Sunil~K.~Mishra\inst{1}}
\shortauthor{H. Verma \etal}
\institute{                    
  \inst{1} Department of Physics, Indian Institute of Technology (Banaras Hindu University), Varanasi - 221005, India\\
  \inst{2} Institut f\"ur Physik, Martin-Luther-Universit\"at Halle-Wittenberg, D-06099 Halle, Germany}
\abstract{
 Qubit(s) transfer through a helical chain is studied. We consider the transfer of a single state and
Bell states across a multiferroic spin chain and the possibility of an 
electric field control of the fidelity of the single state and the  Bell pairs. We analyze pure and imperfect multiferroic spin chains.
A scheme for an efficient transfer of spin states through a multiferroic channel relies on  kicking by appropriate electric field pulses at regular
interval. This electric field pulse sequence undermines the effect of impurity on the fidelity and improves the state transfer through the
helical chain.}
\begin{document}
\newcommand{\newc}{\newcommand}
\newc{\beq}{\begin{equation}}
\newc{\eeq}{\end{equation}}
\newc{\beqa}{\begin{eqnarray}}
\newc{\eeqa}{\end{eqnarray}}
\newc{\kt}{\rangle}
\newc{\br}{\langle}
\newc{\pr}{\prime}
\newc{\longra}{\longrightarrow}
\newc{\ot}{\otimes}
\newc{\rarrow}{\rightarrow}
\newc{\h}{\hat}
\newc{\bom}{\boldmath}
\newc{\btd}{\bigtriangledown}
\newc{\al}{\alpha}
\newc{\be}{\beta}
\newc{\ld}{\lambda}
\newc{\sg}{\sigma}
\newc{\p}{\psi}
\newc{\eps}{\epsilon}
\newc{\om}{\omega}
\newc{\mb}{\mbox}
\newc{\tm}{\times}
\newc{\hu}{\hat{u}}
\newc{\hv}{\hat{v}}
\newc{\hk}{\hat{K}}
\newc{\ra}{\rightarrow}
\newc{\non}{\nonumber}
\newc{\ul}{\underline}
\newc{\hs}{\hspace}
\newc{\longla}{\longleftarrow}
\newc{\ts}{\textstyle}
\newc{\f}{\frac}
\newc{\df}{\dfrac}
\newc{\ovl}{\overline}
\newc{\bc}{\begin{center}}
\newc{\ec}{\end{center}}
\newc{\dg}{\dagger}
\newc{\prh}{\mbox{PR}_H}
\newc{\prq}{\mbox{PR}_q}
\newc{\pd}{\partial}
\newc{\qv}{\vec{q}}
\newc{\dqv}{\delta\vec{q}}
\newc{\dpv}{\delta\vec{p}}
\newc{\mbq}{\mathbf{q}}
\newc{\mbqp}{\mathbf{q'}}
\newc{\mbpp}{\mathbf{p'}}
\newc{\mbp}{\mathbf{p}}
\newc{\mbn}{\mathbf{\nabla}}
\newc{\dmbq}{\delta \mbq}
\newc{\dmbp}{\delta \mbp}
\newc{\T}{\mathsf{T}}
\newc{\J}{\mathsf{J}}
\newc{\sfL}{\mathsf{L}}
\newc{\C}{\mathsf{C}}
\newc{\B}{\mathsf{M}}
\newc{\V}{\mathsf{V}}
\maketitle
\section{Introduction}
Spin chains have long been studied as a credible contender
for carrying out quantum
information processing and
transmission\cite{s1,Osterloh,Vedral,Verstraete,Greiner,Cirac,Lewenstein,s4,Vekua,Chotorlishvili,Sekania}.
Various new experiments have been carried out and numerous
models proposed for spin chain systems \cite{Wiesendanger,Zheludev,Loidl}. The systems of interest in the present paper
is a multiferroic spin chain through which, we seek to
transfer quantum information or qubits. Multiferroic systems  possess intrinsically  coupled magnetic and ferroelectric order parameters \cite{s5, s6, s7, s8}. Hence,
the strong magneto-electric coupling can be utilized as a tool in quantum information processing.
 For spin-driven emergence of ferroelectric polarization,   the ferroelectric order parameter is directly related to the non-collinear
magnetic order. The ferroelectric polarization vanishes in the collinear magnetic  (ferro or antiferromagnetic) phase, 
while in case of a chiral spin order a net polarization remains that can couple
 to an external electric field allowing so for an electric-field control of the magnetic order. 
 Under  a certain geometry of the system and  the applied external fields, magneto-electric coupling term  mimics 
 dynamical Dzyaloshinskii-Moriya interaction. Breaking of the inversion symmetry associated with the Dzyaloshinskii-Moriya 
 interaction may have key consequences for the transfer of quantum information.
In order to identify the influence of electric field we first investigate
the transmission of qubits through a multiferrroic chain with a static constant
electric field and then introduce the electric field kicks at a regular interval. It is noteworthy that a continuous application of the electric 
field with changing the amplitude leads to a complex nature of the time evolution operator {\it i.e.}, of an integral form due to the non commutation of the exchange interaction and the Dzyaloshinskii-Moriya interaction terms in the Hamiltonian.

Realistic systems always have a defect (pinning centers) and/or  embedded impurities. 
Therefore, the study of the pinning centers and the embedded impurities is not only of an academic but also of a practical interest.
To be more realistic we consider the effect of doping the spin chain {\it i.e.}, introducing impurities at specific sites and constructing various new models considering the
types of impurity. Embedded impurities locally modify the strength of the exchange interaction and break the translational invariance.
A naive guess is that the  impurity or the pining center embedded in the system  hinders the propagation of the excitation through the chain. 
We find, as far as the quantum state transfer fidelity is concerned, the picture is not trivial. 
Studying  the possibility of  the transmission of a single qubit through
the impurity-embedded spin chain with realistic  material parameters we identify case  of a better transmission of the qubit as compared to a
pure spin chain. Some particular recipes for  an effective transfer of qubits are analyzed. 

Initially, the system is prepared in fully polarized state (all spins down).
The qubit is injected to the first site of the spin chain and
received at the last site. The parameter for good transmission named "Fidelity"
is discussed later and the expression is calculated elsewhere given these initial conditions and information injection approach.
Such a transmission has practical applications insofar as  various quantum communication protocols require 
the sender and the receiver to share one qubit each of a Bell pair \cite{s2}. We also explore the possibility of transfer of
Bell pair directly by injecting it to the first two qubits of the spin chain and subsequently receiving them at the last and second last sites of
the chain.  The transmission channel in our case, of course, is an open chain.

The paper is organized as follows: Firstly, we specify the model and its hamiltonian. Next, we provide analytical expressions of the quantum state transfer fidelity
for a single qubit and for Bell states. Lastly, we discuss the results of numerical calculations before summarizing the study.

\section{Models}
The Hamiltonian of the multiferroic system reads \cite {Sekania}
\begin{eqnarray}
\mathcal{H}&=&-J_{1}\sum_{i}\vec{S}_{i}.\vec{S}_{i+1}-J_{2}\sum_{i}\vec{S}_{i}.\vec{S}_{i+2}
+B\sum_{i}\vec{S}_i^z
\nonumber \\
&+&\mathcal{E}(t)g_{ME}\sum_{i}{(\vec{S}_{i}\times \vec{S}_{i+1})}^z.
\end{eqnarray}
The constants $J_1$ and $J_2$ characterize nearest and
next-nearest neighbour interaction strengths. Taking competing nearest ferromagnetic $J_1>0$ and next nearest ferromagnetic $J_2<0$ antiferromagnetic interactions
lead to a spin frustration and a non-collinear spin order. $B$ is the magnetic field and $ E$ is the electric field coupled to the ferroelectric polarization. $g_{ME}$
is the magnetoelectric coupling strength. The time varying electric field affects the electric polarization $\vec{P}$ in a manner such that 
$-\vec{\mathcal{E}}(t).\vec{P}=\mathcal{E}(t)g_{ME}\sum_{i}{(\vec{S}_{i}\times \vec{S}_{i+1})}^z$. Here ${(\vec{S}_{i}\times \vec{S}_{i+1})}^z$ is the z- component of the vector chirality. The time dependent electric field has two components namely static field $E_0$ at times between the kicks and $E_1$ at $t=n\tau$ i.e. at the kick time. The embedded impurity is described by
an additional term and the corresponding sites are excluded
from the above Hamiltonian according to the subscribed impurity model. We adopt open boundary conditions in all the cases discussed in the manuscript. We consider the following cases in the manuscript:\\

\begin{figure}[t!]
\centering
\vspace{-2 cm}
\includegraphics[width=1.0\linewidth]{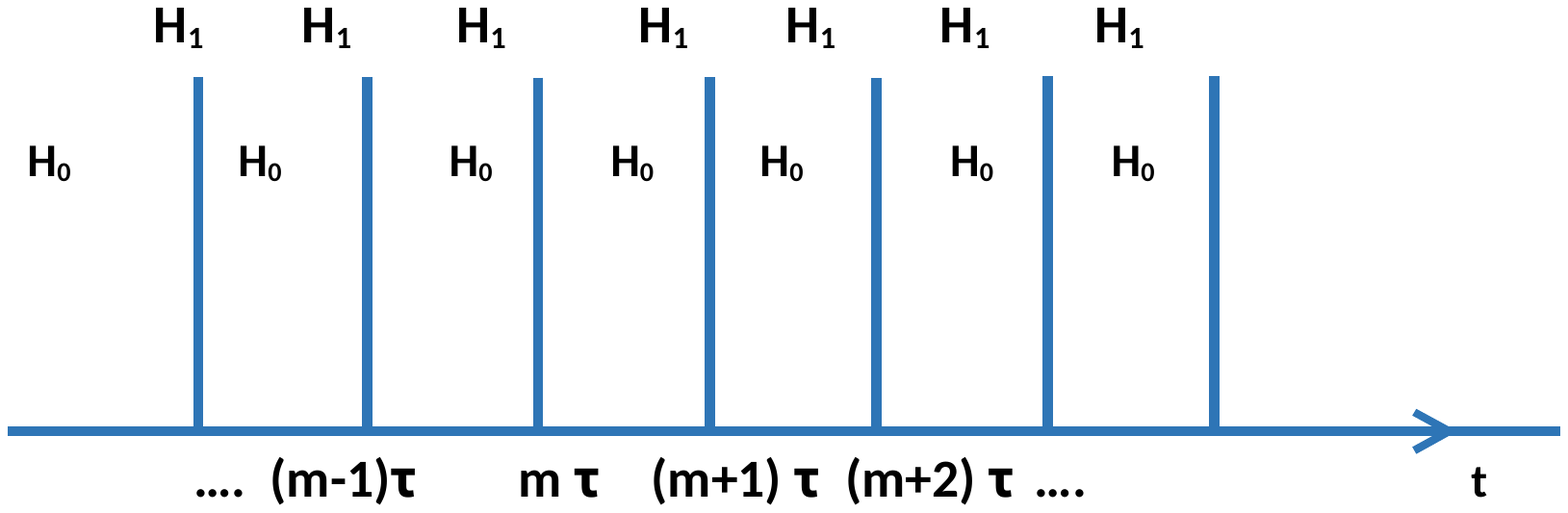}
\vspace{-9cm}
\caption{Kicking scheme showing that the electric field is switched on at a certain regular time interval and can be modelled by  delta functions.}
\label{fig1}
\end{figure}

\textbf	{I. }\textit{Pure Spin Chain with Kicked Electric Field:} The Hamiltonian considered is the above but with
a peculiar twist: Apart from a small static electric field which is switched on at all times, the system is also kicked with an electric field pulse at
regular time intervals. The temporal  profile of the kicking electric field can be taken as  delta function, meaning that the actual duration of the electric field pulse is much smaller that the 
characteristic time scale of the system (in our case below this time scale is in the ps regime, which case sub-ps  electric (laser) field pulses are suitable) 
\cite{andrey_phys_rep}.
As discussed in \cite{andrey_phys_rep} such pulses allow to non-perturbative treatment of the non-equilibrium quantum dynamics.
 The kicking scheme is  illustrated  in Fig. \ref{fig1} and will be used for other spin models in this manuscript. 
Effectively, the Hamiltonian has the following form
\begin{eqnarray}
\mathcal{H}&=&\mathcal{H}_0+\mathcal{H}_1\\
\mathcal{H}_0&=&-J_{1}\sum_{i}\vec{S}_{i}.\vec{S}_{i+1}-J_{2}\sum_{i}\vec{S}_{i}.\vec{S}_{i+2}\nonumber \\
&+& E_0\sum_{i}{(\vec{S}_{i}\times \vec{S}_{i+1})}^z\\
\mathcal{H}_1&=&E_1\sum_{n=1}^{n=\infty} \delta \big(t/\tau- n \big)\sum_{i}
{(\vec{S}_{i}\times \vec{S}_{i+1})}^z
\end{eqnarray}
 where $n$ indicates the number of kicks applied, $E_0=g_{ME}\mathcal{E}_0$ the static background field and $E_1=g_{ME}\mathcal{E}_1$ the amplitude of kicked field. Only at the interval of $\tau$ does the electric field kick term assume non zero value and contributes 
to the Hamiltonian. We have confined our studies to the effect of electric field only and hence $B_z=0$ at all times.

The time evolution operator evaluated between $m\tau_+$  and $(m+ 1)\tau_+$
(Here $t_+$ denotes the time just after the kick) is given by $U(m) = (\hat{\mathcal{U}}_1~\hat{\mathcal{U}}_0)^m$  \cite{andrey_phys_rep} where
\begin{eqnarray}
\hat{\mathcal{U}_0}&=&\exp\bigg(iJ_{1}\tau\sum_{i}\vec{S}_{i}.\vec{S}_{i+1}+iJ_{2}\tau\sum_{i}\vec{S}_{i}.\vec{S}_{i+2} \nonumber \\
&-& iE_0\sum_{i}{(\vec{S}_{i}\times \vec{S}_{i+1})}^z \bigg),
\nonumber \\
\hat{\mathcal{U}}_1&=&\exp\bigg(-iE_1 \sum_{i}{(\vec{S}_{i}\times \vec{S}_{i+1})}^z\bigg),
\label{eq:impulse}
\end{eqnarray}
and the state after the $m^{th}$ kick (or at time $t=m\tau_+$) is 
\begin{eqnarray}
 |\psi(t=m\tau_+)\rangle=\big(\hat{\mathcal{U}}_1~\hat{\mathcal{U}}_0\big)^m~|\psi(t=0)\rangle.
\end{eqnarray}
\textbf	{II. }\textit{Type I impurity:} An impurity is introduced at a site (say between $n$ and $n+1$) in the
spin chain with the usual nearest neighbour and the
next nearest neighbour spin exchange terms.
Effectively, we consider it to be at $n+1$ site by modifying the label of the spins. The introduction
of an impurity between the $n, n+1$ and $n+1, n+2$ sites may lead to a local modification of 
the exchange constants $J_1$ and $J_2$ terms for these most affected spins i.e. a few surrounding the impurity. Keeping in mind that 
both nearest and next nearest exchange interactions remain the same with respect to the impurity, to accommodate this arrangement, change in 
the interaction terms between the subsequent neighbours to $n$ and $n+2$ is considered. It is well known and also been experimentally proved that the 
lattice spacing does effect the exchange integral \cite{s9} i.e. $J_1$ and $J_2$. We just consider the final effect without going into the 
microscopic details and the exact relation by which this parameter changes with the lattice spacing. However, the magnitude of increasing/decreasing
the distance between the lattice sites due to the embedded impurities is assumed to be in tune with changes in $J_1$ and $J_2$.
The following can be noted regarding the impact of introducing  such an impurity on the interaction
between the sets of two spins surrounding the impurity as shown in Fig. \ref{fig2} (a) (b) (c) (d):
\begin{figure}[!t]
\centering
\vspace{-0.5cm}
 \includegraphics*[width=0.65\linewidth]{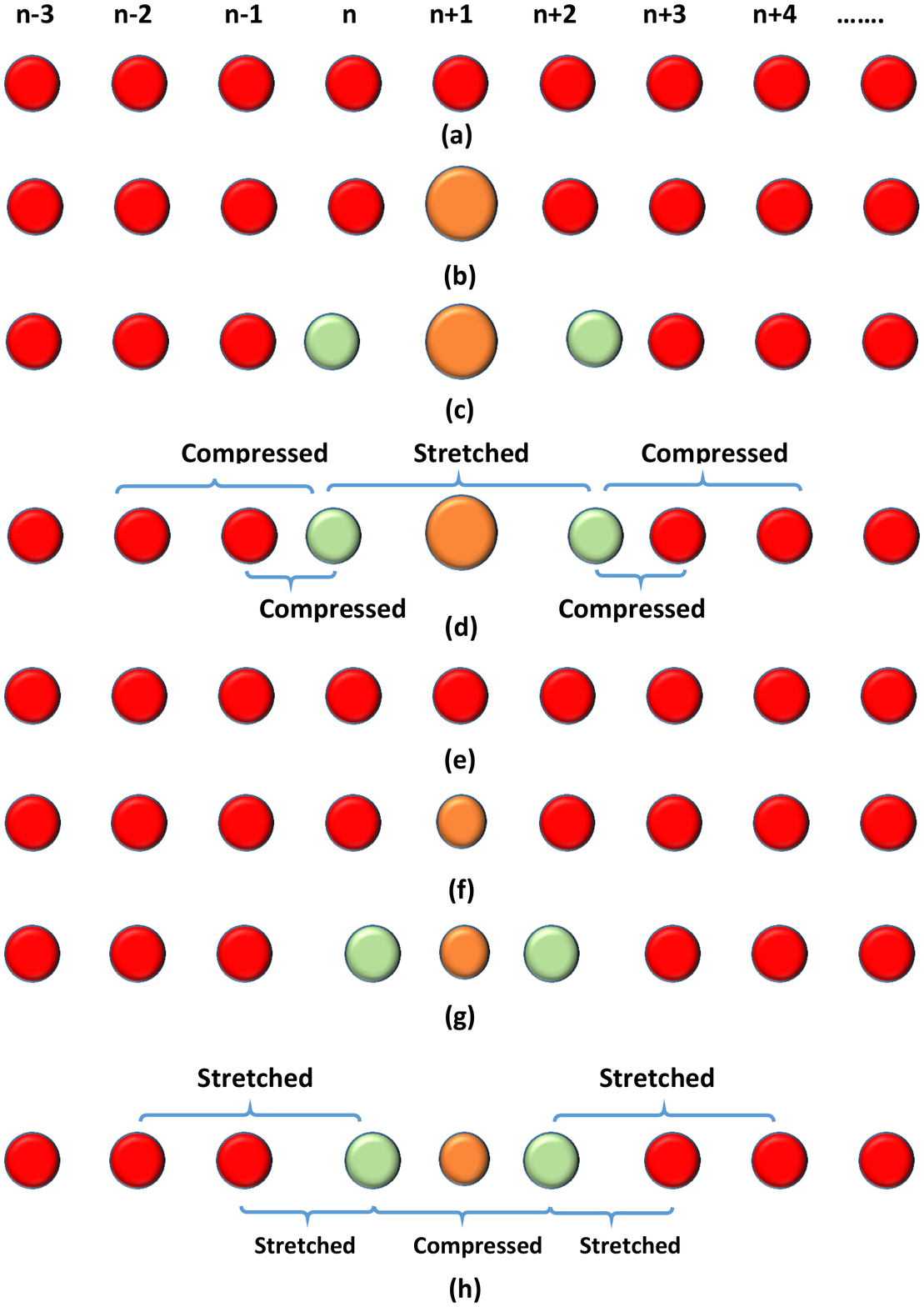}
 \vspace{-1cm}
\caption{The introduction of two types of impurities is depicted. Such impurities cause the local compression or elongation of the bonds between its
nearest and next nearest neighbour. In (a) a pure chain is shown. The impurity
site is chosen as $ n+1$,  as shown in (b). In (c) the rearrangement of spins due to the type I impurity
is shown. The modification in nearest neighbour interaction  ($J_1$) and the next-nearest neighbour interaction ($J_2$) due to this impurity
is shown in (d). Other interactions remain unchanged. (e) Again gives a pure spin chain for reference. Impurity site is shown in (f). (g) Shows the rearrangement of spins due to the introduction of type II impurity. The modifications in nearest neighbour interaction  ($J_1$) and next-nearest neighbour interaction ($J_2$) due to impurity
are shown in (h). Other interactions remain unchanged. }
\label{fig2}
\end{figure}
The nearest neighbour exchange integral ($J_{1}$) are affected: 
[$n-1$, $n$ - increased ($J_{11}$)], [$n$, $n+1$({\it impurity}) - unchanged], [$n+1$({\it impurity}), $n+2$ - unchanged], [$n+2$, $n+3$ - increased ($J_{11}$)].
Next nearest neighbour exchange integral ($J_{2}$) are affected: [$n-2, n$ – increased ($J_{22}$)], [$n-1$, $n+1$({\it impurity}) - unchanged],
[$n$, $n+2$ - decreased ($J_{222}$)],
[$n+1$({\it impurity}), $n+3$ – unchanged], [$n+2$, $n+4$ - increased ($J_{22}$)]. The bonds are being
referenced from the zero impurity model and would correspond to change in
bond parameters ($J_{1}$ or $J_{2}$) i.e. smaller magnitude for stretched bond and
larger magnitude for compressed bond \cite{s9}.
In a nutshell, the impurity changes the interaction
strength in such a way that the nearest and next
nearest bonds of the impurity are unchanged. Other bonds which are
changed are mentioned in Fig. \ref{fig2} and clearly, the effect of introduction of
impurity is localized to 3-4 sites near the impurity.
The Hamiltonian is modified according to the aforementioned scheme to
include the impurity terms.\\

\textbf	{III. }\textit{Type II impurity:} The model is the same as the second one but with just one
change. The embedded impurity locally reduces strength of the exchange interaction between nearest and next nearest neighbouring spins. Similar to the above, we consider impurity to be embedded at
  $n+1$ site by modifying the label of spins. The following can be noted regarding the
  impact of interaction
between sets of two spins surrounding the impurity as shown in Fig. \ref{fig2} (e) (f) (g) (h):

Nearest neighbour exchange integral ($J_{1}$) are affected: [$n-1$, $n$ - decreased ($J_{111}$)], [$n$, $n+1$({\it impurity}) - unchanged],
[$n+1$({\it impurity}), $n+2$ - unchanged], [$n+2$, $n+3$ - decreased ($J_{111}$)]. Next nearest neighbour exchange 
integral ($J_{2}$) affected: [$n-2$, $n$ - decreased ($J_{222}$)],
[$n-1$, $n+1$({\it impurity})- unchanged], [$n$, $n+2$ - increased ($J_{22}$)], [$n+1$({\it impurity}), $n+3$- unchanged], [$n+2$, $n+4$ - decreased ($J_{222}$)].
The bond parameters are changed in the same way as discussed in the previous model.

\section{Fidelity}
\begin{figure}[!t]
 \includegraphics*[trim=0cm 25.5cm 2cm 2cm,width=1\linewidth]{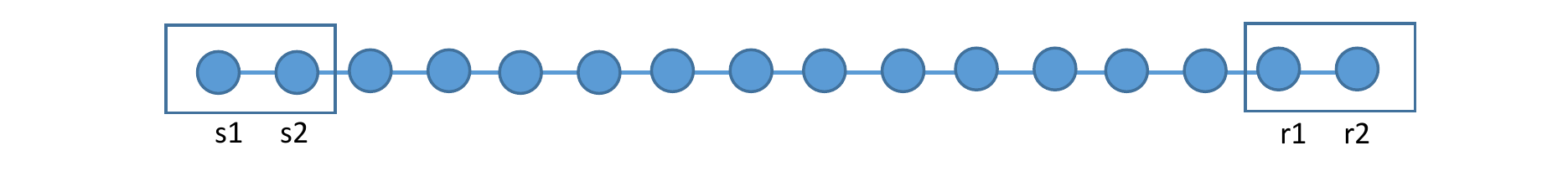}
\caption{Spin channel for transmission of qubit(s). The first two (s1,s2) (one) sites is substituted with Bell pair (single qubit) and 
correspondingly received on the other side at sites r1, r2 (or simply r2).}
\label{fig3}
\end{figure}

For the purpose of transmission of qubit, fidelity may be used as a figure of merit.
Note that as a case in all the models considered, a single spin is substituted at one
end of the chain (sender) and
the qubit is expected to be transmitted to the other end of the chain (receiver). Also, initially the system is prepared in fully spin down polarized state.
The fidelity of qubit transfer is given by \cite{s1}:
$F= \langle\psi_{in}\vert\rho_{out}\vert\psi_{in}\rangle$ which on calculation for a single qubit ($\Omega_0$)
comes out
to be $ F=\frac{\vert f_{r,s}(t_{0})\vert \cos{\gamma}}{3} + \frac{{\vert f_{r,s}(t_{0})\vert}^{2}}{6}+\frac{1}{2} $
where $\gamma= arg(f_{r,s}(t_{0}))$ and $f_{r,s}(t)=\langle r\vert\exp{-iHt}\vert s\rangle$ where
$r$ and $s$ are receiver and sender states respectively and the
Hamiltonian corresponds to a single excitation. A qubit transmitted
through a classical channel has a fidelity of 0.66\cite{s1}. Hence, our interest
lies in conditions for the above systems to exhibit high fidelity, at least greater
than 0.66. Another case in consideration is the transmission
of two qubits through the chain with a protocol very similar to the first case,
i.e. the qubits are substituted at the sites $s1, s2$ and are received at $r1, r2$ (Fig. \ref{fig3}). However, we consider only
the case where the maximally entangled Bell pair is transported.
The fidelity is given by \cite{s3}:
\begin{eqnarray}
\bar{F}(t)&=&\frac{1}{3}\left(| f_{N-1,1}|^2+| f_{N,2}|^2+\frac{| f_{N-1,2}|^2}{2}+\frac{| f_{N,1}|^2}{2}\right)\nonumber \\
&+&\frac{1}{3}Re\left[f_{N,2}f^*_{N-1,1}\right]
\end{eqnarray}
 for the Bell pair $|\Omega_1\kt_{12}=b|01\kt+c|10\kt$, and
\begin{eqnarray}
\bar{F}(t)&=&\frac{1}{2}-\frac{1}{6}\sum_{n=1}^{N-2}\left(| g_{1,2}^{n,N-1}|^2+| g_{1,2}^{n,N}|^2\right)\nonumber\\&+&\frac{1}{3}\left(| g_{1,2}^{N-1,N}|^2+Re\left|g_{1,2}^{N-1,N}\right|\right)
\end{eqnarray}
for $|\Omega_2\kt_{12}=a|00\kt+d|11\kt$ \cite{s3}. 
 Here, $g_{a_1,a_2}^{b_1,b_2}= \br b1,b2\vert\exp{-iHt}\vert a1,a2\kt$. In  this case, the
 subspace of the Hamiltonian corresponds to two spin up excitations among all other spins in down polarized state.
\section{Results}
\begin{figure*}[!t]
\includegraphics*[trim=0cm 8.5cm 0cm 6.5cm, width=1.0\textwidth]{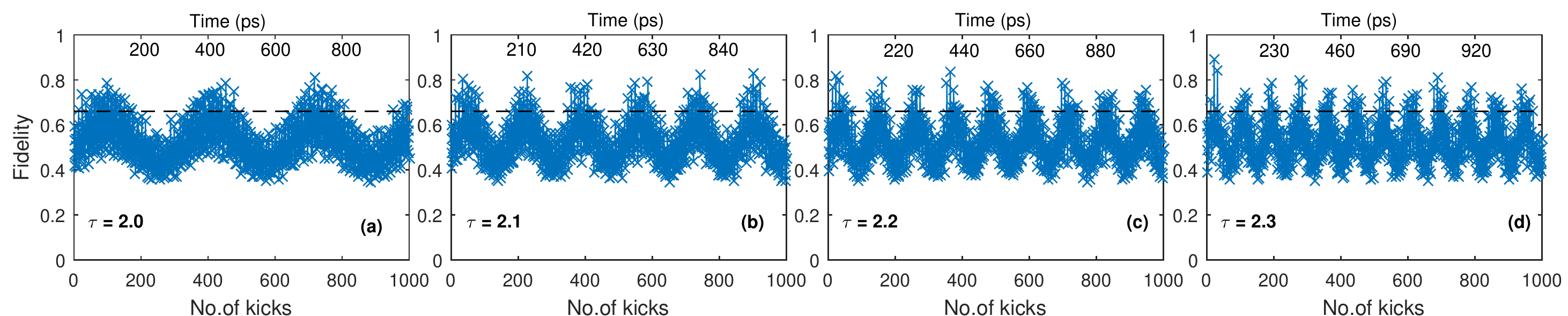} 
\caption{ Fidelity of single state transfer versus number of kicks for kick
intervals $\tau$: (a) 2.0  (b) 2.1 (c) 2.2 (d) 2.3 for $N=10$ sites,
$E_1=1$, $J_1=1$, $J_2=-1$ and $E_0=0.1$. The time axis is taken such that $\tau=1$ corresponds to 0.5 ps which is the kick time scale considered in this paper. The graphs clearly indicate the
periodicity of fidelity observed and as it changes with kick interval. The black dashed line indicates the fidelity through classical channel i.e 0.66 for the single qubit transfer which has been considered.}
\label{fig4}
\end{figure*}

Firstly, the general characteristics of fidelity of a single state transfer and correlation with varying time 
interval between the kicks is shown in Fig. \ref{fig4}. 
The fidelity of a single state transfer shows a distinct periodic behaviour and the control of kick interval gives very visible
results in some region of kick interval (2.0 to 2.3) as shown in Fig. \ref{fig4}.
  
Therefore, it is proposed that by changing the interval between the kicks, the periodicity of fidelity can be changed (and also destroyed). This
emergence of periodicity helps us limit the number of kicks that are sufficient for obtaining the maximum fidelity for spin chains. The increment in 
number of kicks beyond a certain value is of no use as the fidelity pattern repeats. The frequency of this periodic 
form can be ascertained from the Fourier transform. Now that we have an insight into the behaviour of fidelity with a control parameter, {\it i.e.}, time 
interval between the kicks, we move to ascertain the spin chain characteristics for efficient transmission of qubit(s) which is the 
central idea of the paper.

\begin{figure*}[!h]
\begin{center}
 \includegraphics*[trim=0.5cm 8cm 0.5cm 6cm, width=1\linewidth]{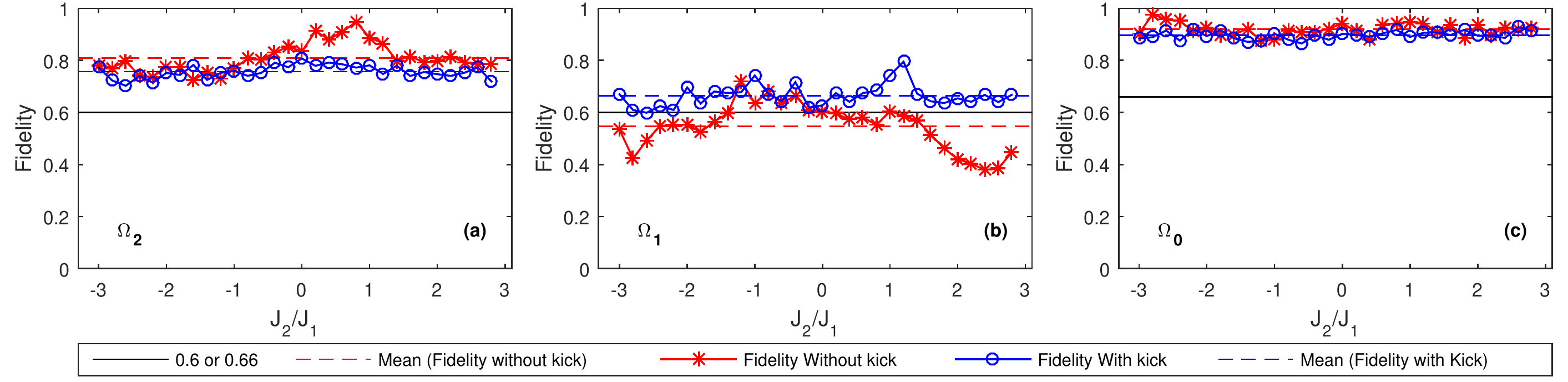}
 \end{center}
 \caption{(a) Fidelity of  $\Omega_2$ state transfer with $E_1=0$ (red) and $E_1=1$ (blue) (b) Fidelity of $\Omega_1$ state transfer with $E_1=0$ (red) and $E_1=1$ (blue)
 (c) Fidelity of single state transfer with $E_1=0$ (red) and $E_1=1$ (blue).
  All chains with parameter $J_2/J_1$ for $N=10$, $J_{1}=1$ and $E_0=0.1$. Kicked electric field is seen to affect the fidelity.}
\label{fig5}
  \end{figure*}

The chains have their maximum fidelity shown in all cases subject to
variable kick interval $\tau$ (0.1 to 10) and
also the number of kicks (up to 500). Without the kicked field, the chain is time evolved through the time interval ranging from 1 to 5000 in tune 
with the maximum time evolution considered for the kicked chain ($500 \times 10$). These parameters are
controllable in a physical system and hence, may be tuned to achieve high
fidelity as obtained in the simulation and the chain engineered to achieve the maximum fidelity
(characterized by $J_2/J_1$ ratio or impurity strength).
For the first case, we compare the results of a spin chain with kicked electric field and without electric field as shown in Fig.~\ref{fig5}.

Clearly, the electric field has a subtle positive effect on the fidelity of single state
transfer for some chains and also shows a normalizing behaviour as seen in Fig. \ref{fig5}(c).
It pushes up the otherwise low in fidelity chains
(characterized by a value of $J_{2}/J_{1}$) and diminishes the
difference of fidelity in different chains at the same time. However, some cases
are found to have been negatively affected by electric field kicks for single state ($\Omega_0$) transfer. 

\begin{figure*}[!h]
 \includegraphics*[trim=0.5cm 9cm 0.5cm 6cm, width=1.0\linewidth]{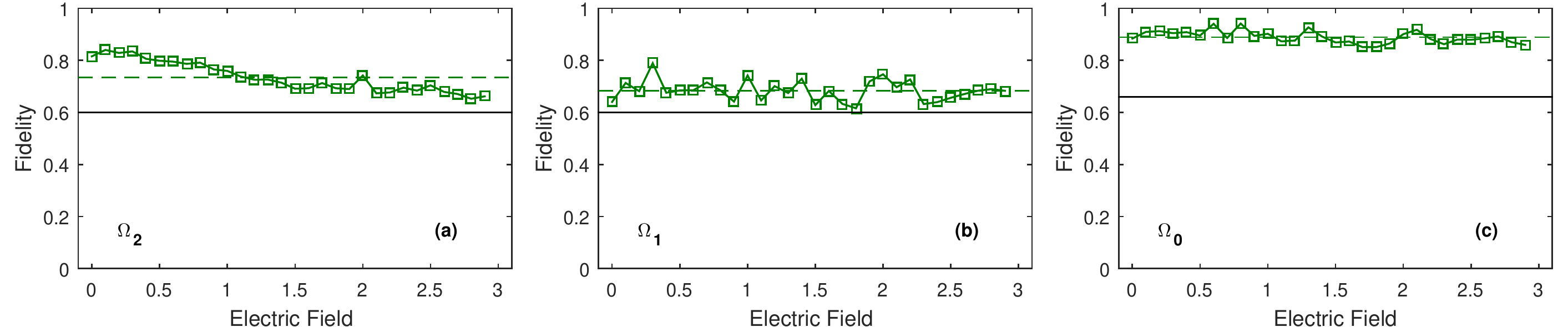}
\caption{(a) Fidelity of $\Omega_2$ state transfer (b) Fidelity of $\Omega_1$ state transfer (c) Fidelity of single state transfer.
 All chains with parameter $E_1$ on x-axis and $J_2/J_1=-1$ for $N=10$, $J_{1}=1$, $E_0=0.1$. Kicked electric field is seen to effect the fidelity.}
\label{fig6}
 \end{figure*}

Now we analyse the fidelity of transfer of two sets of Bell pairs in the
same way as described above from Fig. \ref{fig5} (a), (b).
It is observed that using the electric field kicks improve the fidelity in case of $\Omega_1$ particularly.
Large fidelity enhancement is obtained (reflected in the mean line position) and the
normalizing behaviour is also seen.
The state $\Omega_2$ is negatively effected by the kick subtly but shows the normalizing behaviour.
However, the $\Omega_1$ state is poorly
transferable in such a quantum channel with respect to a classical one albeit few peaks ($J_2/J_1 = -1\ {\rm to} \ 0.8$) are obtained beyond this limit 
even without the kicked electric field, therefore, remains the worst amongst all states despite the enhancement in fidelity due to kicking. Chains
characterized by $J_2/J_1=-2$ to $3$ (with some exceptions) are suitable for transfer of all the
states as they are only ones better than classical (Fig. \ref{fig5}(b)) or almost
the same for transfer of $\Omega_1$ which exhibits least fidelity for transfer among all
the states. We now select the ratio $J_2/J_1$ by seeing its performance in
other input states. It is readily seen from Fig. \ref{fig5}(a) that chains
characterized by $J_2/J_1= -1.6 \ {\rm to} \ 0.8$ are best for $\Omega_2$ state as well. Now, turning to
single state transfer (Fig. \ref{fig5}(c)), we see $J_2/J_1= -1$ to be the
best for our cause exhibiting a fidelity of $>0.9$, although many other chains also suffice our purpose. We are particularly interested in 
the case $J_2/J_1=-1$ as a real material $\rm{LiCu_2O_2}$ having approximately the same parameters can be tested experimentally. Now the electric field is
changed for the selected chain as shown in Fig. \ref{fig6}.

Except some deviations, it can be readily observed that as the kicked
electric field is increased, the fidelity generally decreases for $\Omega_2$ and increases for $\Omega_1$ whereas single state transfer fidelity remains largely unaffected (Fig. \ref{fig6}).
Therefore, the optimum electric fields can be selected again by the peaks of $\Omega_1$ graph as it exhibits the minimum fidelity among all states considered. These are found to be $0.3$, $1.0$, $1.4$. When considered for all the states $E_1=1$ is found to be the most suitable. Interestingly, in Fig. \ref{fig5}, we considered $E_1=1$ which gives the optimum fidelity as found.

\begin{figure*}[!h]
 \includegraphics*[trim=0.5cm 8cm 0.5cm 6cm, width=1.0\linewidth]{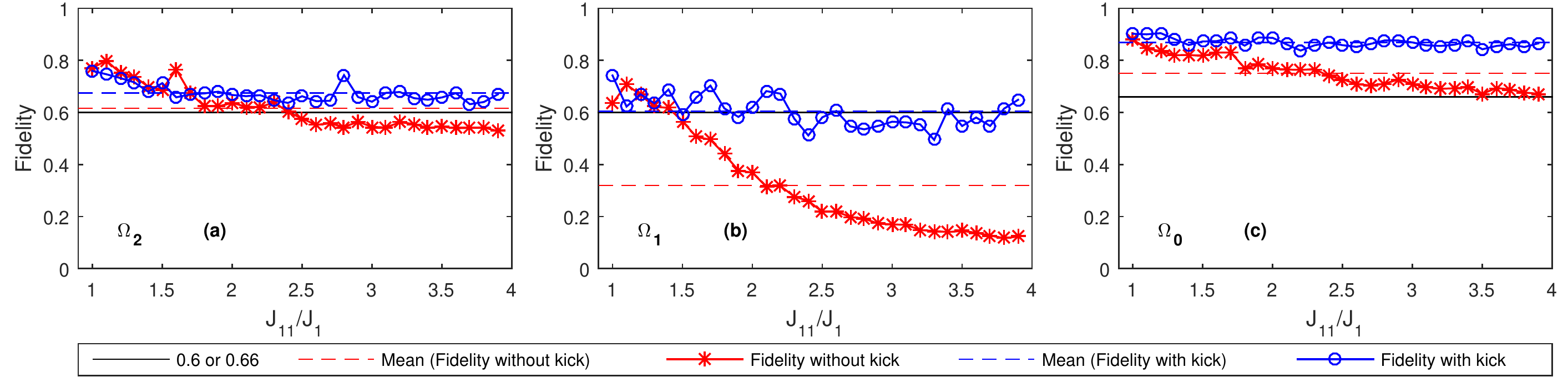}
 \caption{(a) Fidelity of  $\Omega_2$ state transfer with $E_1=0$ (red) and $E_1=1$ (blue) (b) Fidelity of $\Omega_1$ state transfer with $E_1=0$ (red) 
 and $E_1=1$ (blue) (c) Fidelity of $\Omega_0$ state transfer with $E_1=0$ (red) and $E_1=1$ (blue)
  All chains taken with parameter larger nearest neighbour exchange integral $J_{11}$ in chain as shown in Fig. \ref{fig2}. The ratio $J_{11}/J_1$ which is indicative of bond size changes (increases), similarly $J_{22}/J_2$ (next nearest neighbour exchange ratio) and $J_{222}/J_2$ (next nearest neighbour exchange ratio) also change as shown in Fig. \ref{fig2}. $N=10$, $J_{1}=1$, $J_{2}=-1$, $E_0=0.1$ and $E_1=1$. Kicked electric field is seen to effect fidelity.}
  \label{fig7}
\end{figure*}

We now assess the impact of introduction of an impurity showing distinct compression between nearest and next nearest spins (Type I) with the
exchange integral affected as discussed in the previous section. The variation of fidelity with increasing compression is shown in Fig. \ref{fig7}. $J_{11}/J_1$ increases as the effect of impurity is increased and the compression also increases. Starting from $J_{11}/J_1 = J_{22}/J_2 = J_{222}/J_2 =1$, the parameters are slowly changed to simulate increment in the compression and hence the strength of impurity.
We have resorted to $\rm{LiCu_2O_2}$ as the base material in which we embed the impurity and for which, $J_1=11\pm 3$ meV and $J_2=-7\pm 1$ meV and
hence, we can approximately assume $J_1=1$ and $J_2=-1$ in appropriate energy scale. Correspondingly, the energy scale is in the ps regime, meaning 
that the electric-field pulses have to be in the sub-ps time scale for the impulsive approximation in Eq.~\ref{eq:impulse} to be valid.

\begin{figure*}[!t]
\includegraphics*[trim=0.5cm 8cm 0.5cm 6cm, width=1.0\linewidth]{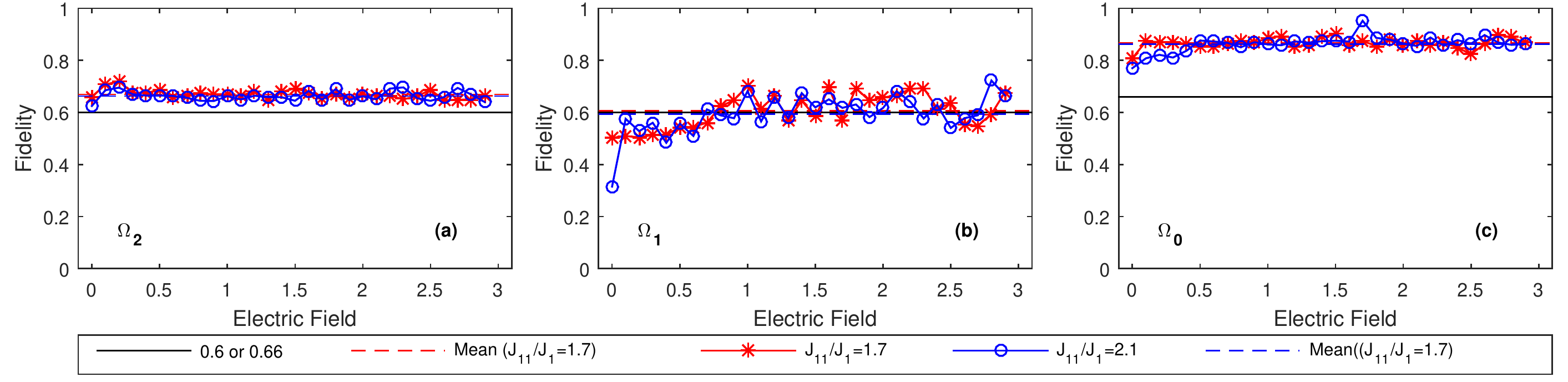}
\caption{(a) Fidelity of $\Omega_2$ state transfer (b) Fidelity of $\Omega_1$ state transfer
(c) Fidelity of single state transfer
  In all the cases the parameter is $E_1$ on x-axis and $J_{11}/J_1=1.7$, $J_{22}/J_2=1.7$ and
  $J_{222}/J_2=0.825$ for red graph and  $J_{11}/J_1=2.1$, $J_{22}/J_2=2.1$ and
  $J_{222}/J_2=0.725$ for the blue graph. $N=10$ , $J_{1}=1$ $J_2=-1$ and $E_0=0.1$. Kicked electric field is seen to effect fidelity.}
  \label{fig8}
\end{figure*}

\begin{figure*}[!h]
\includegraphics*[trim=0.5cm 8cm 0.5cm 6cm, width=1.0\linewidth]{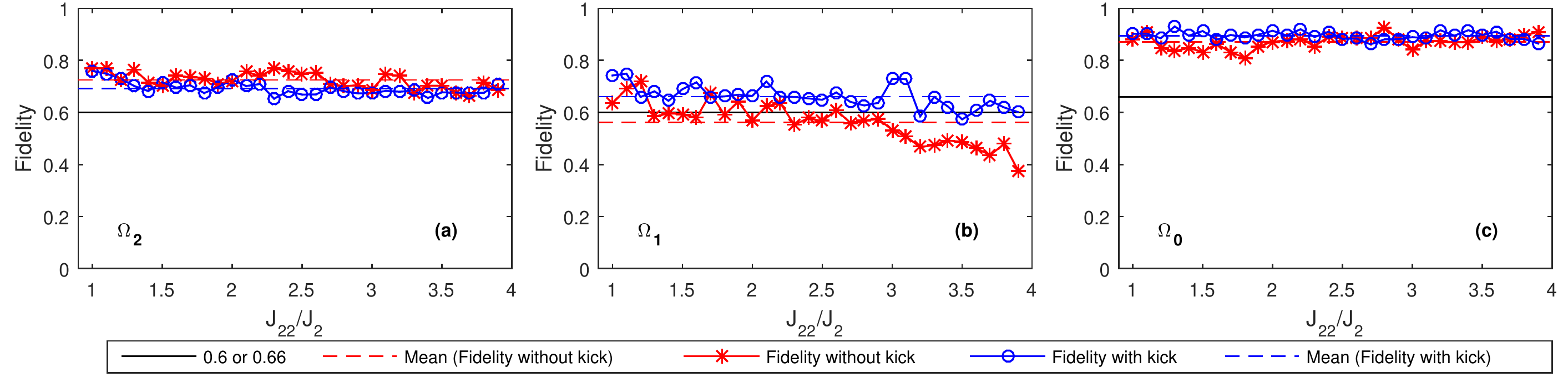}
 \caption{(a) Fidelity of  $\Omega_2$ state transfer with $E_1=0$ (red) and $E_1=1$ (blue) (b) Fidelity of $\Omega_1$ state transfer with $E_1=0$ (red) and $E_1=1$ (blue) (c) Fidelity of single state transfer with $E_1=0$ (red) and $E_1=1$ (blue).
  All chains taken with  shorter next nearest neighbour bond i.e. larger next nearest neighbour exchange integral $J_{22}$ in chain as shown in Fig. \ref{fig2}. The ratio $J_{22}/J_2$ which is indicative of bond size changes, similarly $J_{111}/J_1$ (next nearest neighbour exchange) and $J_{222}/J_2$ (longer next nearest neighbour exchange) also change as shown in Fig. \ref{fig2}. $N=10$, $J_{1}=1$, $J_{2}=-1$ and $E_0=0.1$. Kicked electric field is seen to effect fidelity.}
  \label{fig9}
\end{figure*}

\begin{figure*}[!h]
\includegraphics*[trim=0.5cm 8cm 0.5cm 6cm, width=1.0\linewidth]{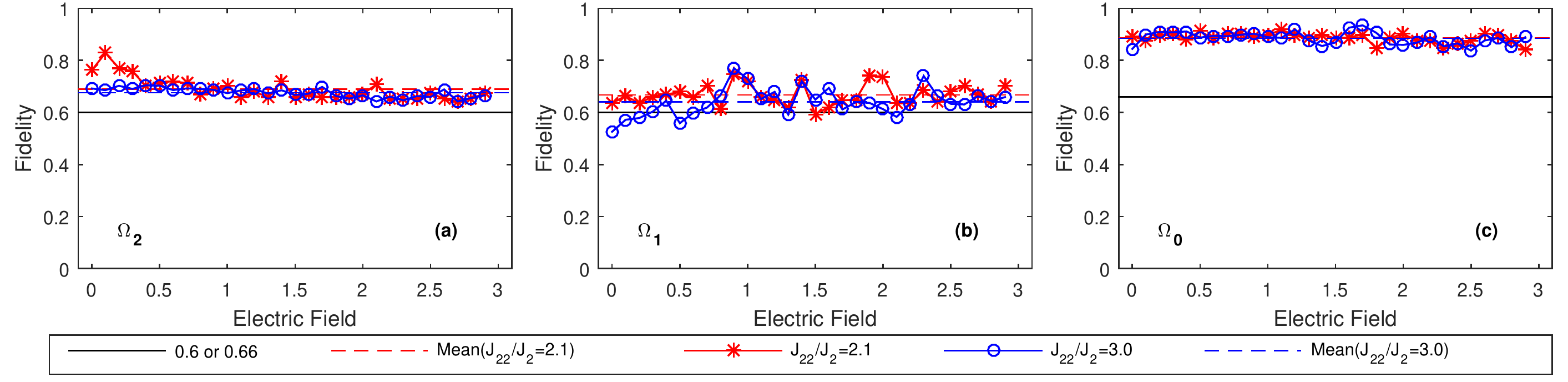}
\caption{(a) Fidelity of $\Omega_2$ state transfer (b) Fidelity of $\Omega_1$ state transfer (c) Fidelity of a single state transfer.
  In all cases, the  parameter is $E_1$ on the x-axis and $J_{22}/J_2=2.1$, $J_{222}/J_2=0.725$ and $J_{111}/J_1=0.725$ for red graph 
  and $J_{22}/J_2=3.0$, $J_{222}/J_2=0.50$ and $J_{111}/J_1=0.50$ for the blue graph. $N=10$ , $J_{1}=1$ $J_2=-1$ and $E_0=0.1$. Kicked electric field is seen to effect fidelity.}
 \label{fig10}
\end{figure*}
There is a marked decrease in the fidelity of chains by the introduction of such an impurity and the increment in compression (strength) decreases
the fidelity for all input states considered.
For the case of two qubit transfer, the already low fidelity for $\Omega_1$ is very low rendering it virtually useless for such a transfer especially
in stronger impurity chains except for the on characterized by $J_{11}/J_1=1.2$ without kicking. But after kicking the chains, some other chains 
characterized by $J_{11}/J_1=1.1, 1.2, 1.3, 1.4, 1.6, 1.7, 2.1$ and $2.2$ are found suitable as seen from Fig. \ref{fig7}(b). On comparing the fidelity for all states it is found that $J_{11}/J_1=1.7$ and $2.1$ are the best choice.

The highest fidelity obtained for $\Omega_2$ was through the pure chain and was reduced by the kicking 
(although above 0.60). The electric field draws a kind of normalizing behaviour for all impurity strengths as is evident 
from Fig. \ref{fig7}. In a way, it undermines the presence of impurity effectively for qubit(s) transfer though at an expense of higher fidelity $\Omega_2$ in some regime. This fidelity is lowered and normalized but the enhancement of fidelity for $\Omega_1$ works to our advantage. For single state also, it is seen to be advantageous (Fig. \ref{fig7}(c)).

Hence, if we implement the kicking scheme with a chain characterized by $J_{11}/J_1$ ratio of $1.7$ or $2.1$, we will be successfully transferring all three states with sufficient fidelity (and relatively better than others).
 Now, we investigate the change in maximal fidelity for the selected chain characterized by such an impurity strength with the magnitude of kicked electric field ($E_1$).

Here, an increase in fidelity is observed with increasing the magnitude of kicked electric field for
single state transfer (Fig. \ref{fig8}(c)) and $\Omega_1$ (Fig. \ref{fig8}(b)). However, fidelity of transfer of $\Omega_2$ remains largely unaffected (Fig. \ref{fig8}(a)). Also, the trends for both the spin chains are nearly the same. Again, $\Omega_{1}$ limits the choice of applied electric
field through the kick to $E_1=1, 1.2, 1.4, 1.6$ and $E_1=1.8 \ {\rm to}\ 2.4$ with some exceptions for both the chains. Isolated peaks at $E_1=2.8$ and $E_1=2.9$ are also observed.

We now calculate the fidelity in the
presence of another type of
impurity (Type II) as discussed in introduction section with resulting elongation between nearest and next nearest spins. The exchange integral are assumed to be affected in the same way as discussed in the introductory section. Starting from $J_{22}/J_2 = J_{222}/J_2 = J_{111}/J_1 =1$, the parameters are slowly changed to simulate increase in impurity strength and the corresponding maximum fidelity noted for all input states considered.
We again resort to $\rm{LiCu_2O_2}$ as the base material for which, $J_1=-11\pm 3$ meV and $J_2=7\pm 1$ meV and
hence, we can approximately assume $J_1=1$ and $J_2=-1$ in appropriate scale.

The results are the same as that obtained in last case of Type I impurities
(Fig. \ref{fig7}) with electric field again enhancing and normalizing the fidelity and subtle decrease in fidelity obtained for all input states as the impurity strength leading to local elongation is increased.
Chains characterized by $J_{22}/J_2=1.1,1.2$ and $1.7$ are found to be the good as they allow
transfer of $\Omega_1$ with sufficient fidelity even without kicking.
By analysing Fig. \ref{fig9}, we may well conclude that chains characterized by
$J_{22}/J_2$ ratio of $2.1$, $3.0$ and many others are suitable for transfer in all three input
cases with kicking. Now the selected chains characterized by
$J_{22}/J_2$ ratio of $2.1$ and $3.0$ are tested against varied electric field.

Once again a marked increase in
fidelity is observed for $\Omega_1$ and $\Omega_0$. Fidelity for transfer of  $\Omega_2$ shows subtle decrease with increase in electric field. 
The peaks in $\Omega_1$ as seen in Fig. \ref{fig10} restricts our choice of electric field to
$E_1\geq0.8$ with some exceptions for different chains. We then move on to the  transfer of $\Omega_2$ and $\Omega_0$ for identification
of particular $E_1$ which is better suited for the complete job which is again $E_1\geq0.8$ as not much advantage is offered by selecting a specific field.

\section{Conclusion}
We have shown here the effectiveness of helical multiferroic chains in transferring 
both- single qubit and Bell pairs ($\Omega_1$ and $\Omega_2$) from one end of
the chain to the other end.
Also, we have found some very interesting properties associated with the
helical chains i.e periodicity of fidelity when
subjected to pulsed (kicked) electric fields. By merely changing
the kick interval, the periodicity is changed or even completely lost.
We have also specifically identified the key parameters
associated with efficient transfer of the qubits viz. electric field and
impurity strength for two types of impurity. In the qubit(s) transfer protocol, the
kick interval and number of kicks are also important for
any particular chain to exhibit the desired fidelity. The recipe
for better transfer is straightforward  and
involves 'kicking' the system (Fig. \ref{fig3}) with electric fields at regular
interval (Fig. \ref{fig1}). By carefully doping the actual
material $\rm{LiCu_2O_2}$ with the impurity whose effect on exchange integral to rest of the
chain we have identified in various cases, the chain can be engineered
to ensure high fidelity. At the same time, results also indicate that the kicking
the system by the electric field undermines the presence of impurity
and chains with distinct strengths of impurity show similar fidelity.
This normalizing behaviour of kicked electric field is seen in both types of impurity
considered and for all the input states. Further, the role of kicked electric
field in transmission of qubits was ascertained by testing selected chains
against the variation of kicked electric field. 
\acknowledgments
SKM acknowledges the support of Department of Science and Technology, India, for support under a grant of
the INSPIRE Faculty Fellowship award [IFA-12 PH 22].

\end{document}